\documentclass[prd, reprint, longbibliography, english, sort&compress, showpacs, showkeys]{revtex4-1}
\usepackage[T1]{fontenc}
\usepackage[latin9]{inputenc}
\usepackage{babel}
\usepackage{amsmath}
\usepackage{amssymb}
\usepackage{esint}
\usepackage[unicode=true,pdfusetitle,
 bookmarks=true,bookmarksnumbered=false,bookmarksopen=false,
 breaklinks=false,pdfborder={0 0 1},backref=false,colorlinks=false]
 {hyperref}
\usepackage{breakurl}

\begin{document}

\title{Linear Response Theory for Symmetry Improved Two Particle Irreducible Effective Actions}

\author{Michael~J.~Brown}
\email{michael.brown6@my.jcu.edu.au}
\affiliation{\emph{College of Science and Engineering, James Cook University, Australia}}

\author{Ian~B.~Whittingham}
%\email{ian.whittingham@jcu.edu.au}
\affiliation{\emph{College of Science and Engineering, James Cook University, Australia}}

\author{Daniel~S.~Kosov}
%\email{daniel.kosov@jcu.edu.au}
\affiliation{\emph{College of Science and Engineering, James Cook University, Australia}}

\date{April 29, 2016}

\begin{abstract}
We investigate the linear response of an $O\left(N\right)$ scalar quantum field theory subject to external perturbations using the symmetry improved two particle irreducible effective action (SI-2PIEA) formalism [A. Pilaftsis and D. Teresi, Nucl. Phys. B874, 594 (2013)].
Despite satisfactory equilibrium behavior, we find a number of unphysical effects at the linear response level.
Goldstone boson field fluctuations are over-determined, with the only consistent solution being to set the fluctuations and their driving sources to zero, except for momentum modes where the Higgs and Goldstone self-energies obey a particular relationship.
Also Higgs field fluctuations propagate masslessly, despite the Higgs propagator having the correct mass.
These pathologies are independent of any truncation of the effective action and still exist even if we relax the over-determining Ward identities, so long as the constraint is formulated $O\left(N\right)$-covariantly.
We discuss possible reasons for the apparent incompatibility of the constraints and linear response approximation and possible ways forward.
\end{abstract}


\maketitle
%\tableofcontents{}

\section{Introduction\label{sec:Introduction}}

Quantum field theory is the mathematical language of nature.
Viewed in this light, much of the last century of theoretical physics can be seen as quantum field theory calculations performed in a variety of approximations.
The most fruitful scheme so far is clearly perturbation theory in small couplings, which has seen wide use and great success.
Nevertheless there are important physical situations where naive perturbation theory fails and must be enhanced by partial resummation if it can be used at all.
For example, massless particles in thermal plasmas produce large loop corrections which must be resummed, causing thermal mass generation and non-analyticities in thermodynamic functions.
Similarly, large logarithms can invalidate naive perturbation theory in problems involving disparate length or energy scales.
Resummation of these logarithms leads to the renormalization group running of coupling constants.
Although the need to go beyond naive perturbation theory is clear in many cases, \emph{ad hoc} resummations are problematic because perturbation series are asymptotic in nature \citep{dyson_divergence_1952}.
Systematic resummation schemes are required to guarantee consistency with the original non-perturbative theory.
There are three such schemes which can claim to be widely studied and successful: the renormalization group (RG) \citep{zinn-justin_phase_2007}, large $N$ expansion \citep{moshe_quantum_2003}, and $n$-particle irreducible effective actions ($n$PIEAs) \citep{berges_introduction_2004}.
This work focuses on the latter.

$n$PIEAs are a functional technique that combine the advantages (and some disadvantages) of perturbation theory and variational methods.
Based on a Legendre transform procedure, $n$PIEAs are guaranteed to be equivalent to the original theory and can capture analytic features of the exact theory that are invisible to perturbation theory, and so side-step potential issues with \emph{ad hoc} resummations \citep{brown_two-particle_2015}.
The 1PIEA, developed by Goldstone, Salam and Weinberg \citep{goldstone_broken_1962} and Jona-Lasinio \citep{jona-lasinio_relativistic_1964}, effects a resummation of tadpoles by defining an effective action functional $\Gamma\left[\varphi\right]$ which depends on the expectation value $\varphi=\left\langle \phi\right\rangle $ of the quantum field.
The physical value of $\varphi$ is determined by the variational equation of motion $\delta\Gamma/\delta\varphi=0$ and, when this equation is satisfied, one can drop all one particle reducible diagrams from the diagrammatic perturbation expansion.

The 2PIEA technique was originally developed by Lee and Yang \citep{lee_many-body_1960}, Luttinger and Ward \citep{luttinger_ground-state_1960}, Baym \citep{baym_self-consistent_1962} and others in the context of many-body theory, then extended by Cornwall,
Jackiw and Tomboulis \citep{cornwall_effective_1974} to relativistic field theory in the functional formalism.
In the 2PIEA formalism one defines an action functional $\Gamma\left[\varphi,\Delta\right]$ depending not just on the mean field, but also its correlation function $\Delta\sim\left\langle \phi\phi\right\rangle -\varphi\varphi$, and solves for $\Delta$ using $\delta\Gamma/\delta\Delta=0$.
This effects a resummation of all self-energy insertions and removes two particle reducible diagrams from the diagrammatic expansion.

Apparently de Dominicis and Martin \citep{de_dominicis_stationary_1964} were the first to realize that the $n=1$ and $n=2$ formalisms are special cases of a general construction giving higher order effective actions for $n\geq3$.
Early work on higher effective actions was carried on by Vasiliev \citep{vasiliev_functional_1998}, whose book was unfortunately not available in English for more than twenty years, though there were some reviews suggesting that the English literature was at least aware of these developments (e.g. \citep{kleinert_higher_1982,haymaker_variational_1991}).
The recent resurgence of interest in higher $n$PIEAs has largely been driven by their advantages for non-equilibrium problems and can likely be credited to the reviews by Berges \citep{berges_introduction_2004,berges_nonequilibrium_2015} and advances in computer power.

One of the disadvantages of $n$PIEAs is that finite order truncations of the effective actions generically do not respect the symmetry properties one expects from the exact theory.
This is because the correlation functions of the 1PIEA obey different Ward identities (WIs) than the 2PIEA etc.
This can also be understood in terms of the patterns of resummations effected by $n$PIEAs which, if truncated, do not preserve the order by order cancellations required to maintain all symmetries.
Similar remarks apply for global symmetries and gauge invariances. The most notable physical consequence of this is a violation of Goldstone's theorem for theories with spontaneous symmetry breaking: would-be Goldstone bosons gain an unphysical mass in the Hartree-Fock approximation, leading also to unphysical reaction thresholds and decay rates.
More subtle effects appear in higher order approximations.
Similarly, in gauge theories a residual gauge dependence appears in quantities that should be physical.
This work focuses on the particular scheme devised by Pilaftsis and Teresi \citep{pilaftsis_symmetry_2013} for improving the symmetry properties of 2PIEAs.
Symmetry improvement enforces 1PI style WIs on the 2PIEA by the use of Lagrange multipliers.
The resulting constraint forces the Goldstone boson correlation functions to obey Goldstone's theorem, giving physically correct massless Goldstone bosons at all orders and even at finite temperatures below the critical temperature of the symmetry breaking phase transition.

So far symmetry improvement has only been applied to non-gauged scalar field theories in equilibrium.
It restores Goldstone's theorem and produces physically reasonable absorptive parts in propagators \citep{pilaftsis_symmetry_2013} and has been shown to restore the second order phase transition of the $O\left(4\right)$ linear sigma model in the Hartree-Fock approximation \citep{mao_symmetry_2014}.
It has been used to study pion strings evolving in the thermal bath of a heavy ion collision \citep{lu_pion_2015} (though note that the symmetry improved 2PIEA was only used to calculate an equilibrium finite temperature effective potential; this work did not constitute what we would call a true non-equilibrium calculation using symmetry improved effective actions).
Symmetry improvement has also been demonstrated to improve the evaluation of the effective potential of the standard model by taming the infrared divergences of the Higgs sector, treated as an $O\left(4\right)$ scalar field theory with gauge interactions turned off \citep{pilaftsis_symmetry_2015,pilaftsis_symmetry-improved_2015}.
The symmetry improvement idea can also be extended to 3PIEAs \citep{brown_symmetry_2015}, which requires enforcing an additional set of WIs for the three point vertex function and the fixing of an ambiguity in the constraint procedure.
During review of this manuscript we became aware of the recent preprint \citep{marko_loss_2016} which shows that symmetry improved 2PI equations of motion can lose solutions even in equilibrium as the result of a truncation artifact.
This loss of solutions is distinct from the one we find below and shows that more work is required to understand the domain of applicability of symmetry improved effective actions.

There is a strong motivation to extend symmetry improvement beyond equilibrium since one of the major reasons for using $n$PIEAs in the first place is their ability to handle non-equilibrium situations.
$n$PIEAs give an entirely mechanical way to set up the generic initial value problem as a \emph{closed} system of integro-differential equations directly for the mean fields and low order correlation functions, which are simply related to the handful of physical quantities (densities,
conserved currents, etc.) one is most often interested in.
Apart from a truncation to some finite loop order these equations need not be subject to any further approximation.
Hence, apart from the symmetry issue and issues involved in the renormalization process, $n$PIEAs give potentially the most general and accurate framework available for the computation of real time properties in quantum field theory.
Motivated by this, we seek to extend the symmetry improvement technique to non-equilibrium situations.
The ultimate goal of this program would be a tractable and manifestly gauge invariant set of equations of motion for highly excited Yang-Mills-Higgs theories with chiral fermion matter based on the self-consistently complete 4PIEA.
In the meantime we content ourselves with an analysis of the symmetry improved 2PIEA for scalar fields in the linear response regime.

We investigate the linear response approximation rather than a generic non-equilibrium situation for several reasons.
First, the linear response approximation is simply far more tractable than the general non-equilibrium situation as the response functions only depend on the equilibrium properties of the theory.
Second, linear response is widely applicable in the real world: many systems are ``close enough'' to equilibrium for practical purposes.
Third, the linear response approximation is a nice laboratory to isolate the novel features of symmetry improvement constraints in non-equilibrium settings.
Indeed, we can find results that are independent of any truncation of the effective action.
Finally, we expect any physically reasonable formalism to give a well formed linear response approximation, though this depends on the assumption that the exact behavior is an analytic (or at least not too singular) function of the external perturbation within some neighborhood of zero perturbation.
This is true of all quantum mechanical systems (so long as the Hamiltonian remains bounded below), but for field theories the infinite number of degrees of freedom may complicate the situation.

The outline of the remainder of this work is as follows.
In Section \ref{sec:Linear-Response-Theory} we review linear response theory and 2PIEAs.
Then in Section \ref{sec:Symmetry-Improvement} we review the symmetry improvement method, re-deriving the WI constraints in the presence of external sources which were neglected in previous work.
In Section \ref{sec:Implications-of-Constraints} we derive the consequences of the constraints for the linear response functions, noting that a careful treatment of the constraint procedure requires that not just the WI, but also its derivatives, must vanish.
In Section \ref{sec:Discussion} we reach our conclusions about the feasibility of symmetry improvement within the linear response approximation and sketch some ideas for future work.
In Appendix \ref{sec:The-Alternate-Constraint} we discuss an alternative symmetry improvement procedure that relaxes the over-determining WIs, but still leads to difficulties due to the necessity of the derivatives of the constraint to also vanish.
Finally, Appendix \ref{sec:Mechanical-Analogy} includes a mechanical analogy which illustrates some of the subtle points about the constraint procedure in a simpler setting.

Our conventions follow our previous paper \citep{brown_symmetry_2015}.
In particular $\hbar=c=1$ and $\eta_{\mu\nu}=\mathrm{diag}\left(1,-1,-1,-1\right)$.
Loop counting factors of $\hbar$ will be kept.
Repeated indices are summed, and spacetime arguments going along with $O\left(N\right)$ indices are implicitly integrated over (``DeWitt notation'').
Where explicitly written spacetime integrals are $\int_{x}\equiv\int\mathrm{d}^{4}x$ and momentum integrals are $\int_{p}=\int\mathrm{d}^{4}p/\left(2\pi\right)^{4}$.
$\left(\tilde{\mathrm{T}}\right)\mathrm{T}\left[\cdots\right]$ represents the (anti-)time ordered product of the factors in $\left[\cdots\right]$.
It is not necessary here to distinguish between real time, Matsubara and Schwinger-Keldysh time contours.
For a physical quantity $X$ we denote its equilibrium value by $\tilde{X}$ and its shift under linear response by $\delta X$ so that $X=\tilde{X}+\delta X$ plus higher order terms.

\section{Linear Response Theory and Effective Actions\label{sec:Linear-Response-Theory}}

Linear response theory studies the effect of small externally applied perturbations on a system initially in equilibrium.
Consider a quantum system which is subjected to an external driving potential $-J\left(t\right)\hat{B}\left(t\right)$ where $J\left(t\right)$ is a c-number function of time representing the strength of the driving and $\hat{B}\left(t\right)$ is the interaction Hamiltonian (the reason for the name will become apparent).
If the initial state of the system is described by a density matrix $\rho_{0}$ at time $t_{0}$, with $J\left(t\right)=0$ for $t\leq t_{0}$, then at time $t>t_{0}$ the expectation of an operator $\hat{A}$ (in the interaction picture with respect to the external perturbation) is:
\begin{align}
\left\langle \hat{A}\left(t\right)\right\rangle  & =\mathrm{Tr}\left\{ \rho\left(t\right)\hat{A}\left(t\right)\right\} \nonumber \\
 & =\mathrm{Tr}\left\{ U\left(t,t_{0}\right)\rho_{0}U\left(t,t_{0}\right)^{\dagger}\hat{A}\left(t\right)\right\} \nonumber \\
 & =\mathrm{Tr}\left\{ \mathrm{T}\left[\mathrm{e}^{i\int_{t_{0}}^{t}J\left(\tau\right)\hat{B}\left(\tau\right)\mathrm{d}\tau}\right]\rho_{0} \right. \nonumber \\
 &    \left. \times \tilde{\mathrm{T}}\left[\mathrm{e}^{-i\int_{t_{0}}^{t}J\left(\tau\right)\hat{B}\left(\tau\right)\mathrm{d}\tau}\right]\hat{A}\left(t\right)\right\} \nonumber \\
 & =\left\langle \tilde{\hat{A}}\left(t\right)\right\rangle+i\int_{t_{0}}^{t}\left\langle \left[\hat{A}\left(t\right),\hat{B}\left(\tau\right)\right]\right\rangle J\left(\tau\right)\mathrm{d}\tau \nonumber \\
&    +\mathcal{O}\left(J^{2}\right),
\end{align}
where $\left\langle \tilde{\hat{A}}\left(t\right)\right\rangle $ denotes the expected value in the absence of perturbation.
This leads us to define the response function $\chi^{AB}\left(t-\tau\right)=i\left\langle \left[\hat{A}\left(t\right),\hat{B}\left(\tau\right)\right]\right\rangle \Theta\left(t-\tau\right)$ (which only depends on the time difference due to the equilibrium assumption about $\rho_{0}$) such that
\begin{equation}
\delta A\left(t\right)\equiv\left\langle \hat{A}\left(t\right)\right\rangle -\left\langle \tilde{\hat{A}}\left(t\right)\right\rangle =\int_{t_{0}}^{t}\chi^{AB}\left(t-\tau\right)J\left(\tau\right)\mathrm{d}\tau+\mathcal{O}\left(J^{2}\right).
\end{equation}
(The limits can be pushed to $\pm\infty$ thanks to the step function in $\chi^{AB}$, and the equation becomes trivial in the Fourier domain.)
The goal of linear response theory is to compute $\chi^{AB}\left(t-\tau\right)$ for perturbations $\hat{B}$ and observables $\hat{A}$ of interest.
The condition for validity of the approximation is that the quadratic term, which is 
\begin{equation}
-\int_{t_{0}}^{t}\int_{t_{0}}^{\tau_{1}}J\left(\tau_{1}\right)J\left(\tau_{2}\right)\left\langle \left[\left[\hat{A}\left(t\right),\hat{B}\left(\tau_{1}\right)\right],\hat{B}\left(\tau_{2}\right)\right]\right\rangle \mathrm{d}\tau_{2}\mathrm{d}\tau_{1},
\end{equation}
is much smaller than the linear term, which occurs for sufficiently small sources $J$ and times $t-t_{0}$.

We now specialize to the scalar $O\left(N\right)$ field theory defined by the action
\begin{equation}
S\left[\phi\right]=\int_{x}\mathcal{L}\left[\phi\right]=\int_{x}\frac{1}{2}\partial_{\mu}\phi_{a}\partial^{\mu}\phi_{a}-\frac{1}{2}m^{2}\phi_{a}\phi_{a}-\frac{\lambda}{4!}\left(\phi_{a}\phi_{a}\right)^{2},
\end{equation}
where $a=1,\cdots,N$ and we choose to operate in the spontaneous symmetry breaking regime with $m^{2}<0$.
The vacuum expectation value $v$ is given by $v^{2}=-6m^{2}/\lambda$ and the tree level mass of the radial (``Higgs'') mode is $m_{H}^{2}=\lambda v^{2}/3$.
We take the vacuum expectation value to be in the last component $\varphi_{a}\equiv\left\langle \phi_{a}\right\rangle =\left(0,\cdots,0,v\right)$.
The symmetry transformation is $\delta\phi_{a}=i\epsilon_{A}T_{ab}^{A}\phi_{b}$ where $T_{ab}^{A}$ with $A=1,\cdots,N\left(N-1\right)/2$ are the generators of rotations.
When we have cause to use specific generators we will write $A=\left(j,k\right)$ with $j\neq k$ in $1,\cdots,N$ to denote the plane of rotation, and have $T_{ab}^{jk}=i\left(\delta_{ja}\delta_{kb}-\delta_{jb}\delta_{ka}\right)$.

Considering this theory in a particle physics context, the most likely external perturbations will be linear or quadratic in the fields.
For example, we could be using this theory to describe mesons coupling to hadrons through Yukawa interactions $\sim\phi\bar{\psi}\psi$, or Higgs fields coupling to a Yang-Mills sector through (possibly some subset of) the conserved currents with terms like $W^{\mu A}\left(i\phi_{a}T_{ab}^{A}\overleftrightarrow{\partial_{\mu}}\phi_{b}\right)$ and $W^{\mu A}W_{\mu A}\phi_{a}\phi_{a}$.
The theory can also represent an extended Higgs or dark matter sector coupled via portal terms to the standard model Higgs with $\sim\phi_{a}\phi_{a}\Phi^{\dagger}\Phi$, or coupled to a standard model singlet scalar sector $S$ through $\sim S\phi\phi$.
Finally the fields could represent a multi-field inflaton, moduli or an extended gravitational sector, all of whose interactions to the standard model sector will be mediated by terms of the preceding forms to leading order in an effective field theory expansion.
Thus, on very general grounds we expect that the external perturbation can be taken as a linear or quadratic function of the fields.
The only notable exception is the coupling of the field theory to gravity, which includes the quartic interaction $\sim\sqrt{-g}\left(\phi_{a}\phi_{a}\right)^{2}$.

We now consider the notable observables in the theory.
Chiefly we will be interested in the field expectation values $\varphi_{a}$, the conserved $O\left(N\right)$ currents
\begin{equation}
j_{\mu}^{A}=i\left\langle \phi_{a}T_{ab}^{A}\overleftrightarrow{\partial_{\mu}}\phi_{b}\right\rangle ,
\end{equation}
and the energy-momentum tensor
\begin{align}
T_{\mu\nu} &=\left\langle \partial_{\mu}\phi_{a}\partial_{\nu}\phi_{a} \vphantom{\left(\frac{1}{2}\right)} \right. \nonumber \\
 & \left. -\eta_{\mu\nu}\left(\frac{1}{2}\partial_{\rho}\phi_{a}\partial^{\rho}\phi_{a} -\frac{1}{2}m^{2}\phi_{a}\phi_{a}-\frac{\lambda}{4!}\left(\phi_{a}\phi_{a}\right)^{2}\right)\right\rangle .
\end{align}
Again we see that, apart from the $\left(\phi_{a}\phi_{a}\right)^{2}$ term in the energy, the observables are also linear or quadratic in the fields.
Note that if desired $\left\langle \left(\phi_{a}\phi_{a}\right)^{2}\right\rangle $ can be approximated in a mean field approximation by terms of the form $\left\langle \phi\phi\right\rangle ^{2}$, $\left\langle \phi\phi\right\rangle \left\langle \phi\right\rangle ^{2}$ and $\left\langle \phi\right\rangle ^{4}$ plus corrections that can be found diagrammatically.
This means that many of the local quantities one might want can be determined by a method yielding $\varphi_{a}$ and $i\hbar\Delta_{ab}\equiv\left\langle \mathrm{T}\left[\phi_{a}\phi_{b}\right]\right\rangle -\varphi_{a}\varphi_{b}$ subject to generic quadratic perturbations of the form $-J_{a}\phi_{a}-\frac{1}{2}K_{ab}\phi_{a}\phi_{b}$.
This problem seems tailor made for the 2PIEA.

The 2PIEA is defined as the double Legendre transform of the connected generating functional
\begin{equation}
W\left[J,K\right]=-i\hbar\ln Z\left[J,K\right],
\end{equation}
where the partition function is
\begin{equation}
Z\left[J,K\right]=\int\mathcal{D}\left[\phi\right]\exp\frac{i}{\hbar}\left(S\left[\phi\right]+J_{a}\phi_{a}+\frac{1}{2}\phi_{a}K_{ab}\phi_{b}\right).
\end{equation}
Here $J_{a}$ and $K_{ab}=K_{ba}$ are the externally applied sources coupling linearly and quadratically to the field respectively.
Performing the double Legendre transform gives the 2PIEA functional
\begin{equation}
\Gamma\left[\varphi,\Delta\right]=W-J_{a}\frac{\delta W}{\delta J_{a}}-K_{ab}\frac{\delta W}{\delta K_{ab}},
\end{equation}
where on the right hand side $J_{a}$ and $K_{ab}$ are eliminated in terms of $\varphi_{a}$ and $\Delta_{ab}$ by inverting
\begin{align}
\frac{\delta W}{\delta J_{a}} & =\varphi_{a},\\
\frac{\delta W}{\delta K_{ab}} & =\frac{1}{2}\left(i\hbar\Delta_{ab}+\varphi_{a}\varphi_{b}\right).
\end{align}
The equations of motion for $\Gamma\left[\varphi,\Delta\right]$ are then
\begin{align}
\frac{\delta\Gamma}{\delta\varphi_{a}} & =-J_{a}-K_{ab}\varphi_{b},\label{eq:unmodified-vev-eom}\\
\frac{\delta\Gamma}{\delta\Delta_{ab}} & =-\frac{1}{2}i\hbar K_{ab}.\label{eq:unmodified-Delta-eom}
\end{align}

The result of performing the Legendre transform is the expression
\begin{equation}
\Gamma\left[\varphi,\Delta\right]=S\left[\varphi\right]+\frac{i\hbar}{2}\mathrm{Tr}\ln\left(\Delta^{-1}\right)+\frac{i\hbar}{2}\mathrm{Tr}\left(\Delta_{0}^{-1}\Delta\right)+\Gamma_{2}\left[\varphi,\Delta\right],\label{eq:2piea}
\end{equation}
where the free propagator is
\begin{align}
\Delta_{0ab}^{-1}\left(x,y\right)&=\frac{\delta S}{\delta\phi_{a}\left(x\right)\delta\phi_{b}\left(y\right)}\left[\varphi\right]\nonumber \\
& =\left[\left(-\square_{x}-m^{2}-\frac{\lambda}{6}\varphi_{c}\varphi^{c}\right)\delta_{ab}-\frac{\lambda}{3}\varphi_{a}\varphi_{b}\right] \nonumber \\
& \times\delta\left(x-y\right),
\end{align}
and $\Gamma_{2}\left[\varphi,\Delta\right]$ is the sum of all two particle irreducible vacuum Feynman diagrams with propagators $\Delta$ and vertices obtained from the cubic and quartic parts of the shifted action $S\left[\phi+\varphi\right]$.
The equation of motion for $\Delta$ (in the absence of sources) is simply the Dyson equation:
\begin{equation}
\Delta^{-1}=\Delta_{0}^{-1}-\Sigma,\label{eq:dyson-eq}
\end{equation}
where the self-energy is
\begin{align}
\Sigma_{ab}\left(x,y\right) & =\frac{2i}{\hbar}\frac{\delta\Gamma_{2}}{\delta\Delta_{ab}\left(x,y\right)}\nonumber \\
 & =i\hbar\frac{\lambda}{6}\left[\delta_{ab}\Delta_{cc}\left(x,x\right)+2\Delta_{ab}\left(x,x\right)\right]\delta\left(x-y\right)\nonumber \\
 & +i\frac{\hbar}{2}\int_{rsuv}V_{acd}\left(x,r,s\right)V_{bef}\left(y,u,v\right) \nonumber \\
 & \times \Delta_{ce}\left(r,u\right)\Delta_{df}\left(s,v\right) +\mathcal{O}\left(\hbar^{2}\right),
\end{align}
where the second line is the expansion to one loop order and we have introduced the three point vertex function
\begin{align}
V_{abc}\left(x,y,z\right)&=-\frac{\lambda}{3}\left[\delta_{ab}\varphi_{c}\left(x\right)+\delta_{ca}\varphi_{b}\left(x\right)+\delta_{bc}\varphi_{a}\left(x\right)\right]\nonumber \\
 & \times \delta\left(x-y\right)\delta\left(x-z\right)
\end{align}
for convenience.
The \emph{Hartree-Fock} approximation is obtained by retaining only the $O\left(\lambda\right)$ term in $\Sigma$.
Note that none of our results actually require any truncation of $\Sigma$.

For the equilibrium solution we make the spontaneous symmetry breaking ansatz
\begin{align}
\tilde{\varphi}_{a} & =v\delta_{aN},\label{eq:eq-ssb-ansatz-vev}\\
\tilde{\Delta}_{ab}\left(x,y\right) & =\begin{cases}
\Delta_{G}\left(x,y\right), & a=b\neq N,\\
\Delta_{H}\left(x,y\right), & a=b=N,\\
0, & \text{otherwise},
\end{cases}\label{eq:eq-ssb-ansatz-delta}
\end{align}
where $\Delta_{G/H}$ are the Goldstone/Higgs propagators respectively.
We define the masses $m_{G/H}^{2}$ from the corresponding exact propagators.
The free propagator is
\begin{eqnarray}
\Delta_{0ab}^{-1}\left(x,y\right) & = & -\left(\square_{x}\delta_{ab}+m_{ab}^{2}\right)\delta\left(x-y\right),\\
m_{ab}^{2} & = & \begin{cases}
m^{2}+\frac{\lambda v^{2}}{6}, & a=b\neq N,\\
m^{2}+\frac{\lambda v^{2}}{6}+\frac{\lambda v^{2}}{3}, & a=b=N,\\
0, & \text{otherwise},
\end{cases}
\end{eqnarray}
so that $m_{G}^{2}=m^{2}+\lambda v^{2}/6$ and $m_{H}^{2}=m_{G}^{2}+\lambda v^{2}/3$ to lowest order.
However, truncations of the vev equation of motion $\delta\Gamma/\delta v=0$ \emph{do not} generically obey $v=-6m^{2}/\lambda$ so that $m_{G}^{2}\neq0$.
Symmetry improvement replaces the vev equation of motion by the constraint $vm_{G}^{2}=0$ so that Goldstone's theorem is exactly satisfied whenever $v\neq0$.
The value of $v$ is then determined by the self-consistent solution of (\ref{eq:dyson-eq}).

To connect this formalism to linear response theory we must expand $\varphi\to\tilde{\varphi}+\delta\varphi$ and $\Delta\to\tilde{\Delta}+\delta\Delta$ about their source-free equilibrium values $\tilde{\varphi}$ and $\tilde{\Delta}$ in (\ref{eq:unmodified-vev-eom})-(\ref{eq:unmodified-Delta-eom}) and match terms order by order in the sources, treating the responses $\delta\varphi$ and $\delta\Delta$ as first order, as typical of a perturbation theory analysis.
At lowest order we find (\ref{eq:unmodified-vev-eom})-(\ref{eq:unmodified-Delta-eom}) with no sources for the equilibrium solutions and at first order we find
\begin{align}
\frac{\delta^{2}\Gamma}{\delta\varphi_{b}\delta\varphi_{a}}\delta\varphi_{b}+\frac{\delta^{2}\Gamma}{\delta\Delta_{bc}\delta\varphi_{a}}\delta\Delta_{bc} & =-J_{a}-K_{ab}\tilde{\varphi}_{b},\\
\frac{\delta^{2}\Gamma}{\delta\varphi_{c}\delta\Delta_{ab}}\delta\varphi_{c}+\frac{\delta^{2}\Gamma}{\delta\Delta_{cd}\delta\Delta_{ab}}\delta\Delta_{cd} & =-\frac{1}{2}i\hbar K_{ab},
\end{align}
where all derivatives on the left hand sides are evaluated at the equilibrium values.
It is possible to eliminate the fluctuations from these equations by introducing the linear response functions $\chi_{ab}^{\phi J}$, $\chi_{abc}^{\phi K}$, $\chi_{abc}^{\Delta J}$ and $\chi_{abcd}^{\Delta K}$:
\begin{align}
\delta\varphi_{a} & =\chi_{ab}^{\phi J}J_{b}+\frac{1}{2}\chi_{abc}^{\phi K}K_{bc},\\
\delta\Delta_{ab} & =\chi_{abc}^{\Delta J}J_{c}+\frac{1}{2}\chi_{abcd}^{\Delta K}K_{cd},
\end{align}
and demanding that the resulting equations hold for any value of the sources $J$, $K$.
Doing this leads to the system
\begin{align}
\frac{\delta^{2}\Gamma}{\delta\varphi_{b}\delta\varphi_{a}}\chi_{bd}^{\phi J}+\frac{\delta^{2}\Gamma}{\delta\Delta_{bc}\delta\varphi_{a}}\chi_{bcd}^{\Delta J} & =-\delta_{ad},\\
\frac{\delta^{2}\Gamma}{\delta\varphi_{c}\delta\Delta_{ab}}\chi_{ce}^{\phi J}+\frac{\delta^{2}\Gamma}{\delta\Delta_{cd}\delta\Delta_{ab}}\chi_{cde}^{\Delta J} & =0,\\
\frac{\delta^{2}\Gamma}{\delta\varphi_{b}\delta\varphi_{a}}\chi_{bde}^{\phi K}+\frac{\delta^{2}\Gamma}{\delta\Delta_{bc}\delta\varphi_{a}}\chi_{bcde}^{\Delta K} & =-\left(\delta_{ad}\tilde{\varphi}_{e}+\delta_{ae}\tilde{\varphi}_{d}\right),\\
\frac{\delta^{2}\Gamma}{\delta\varphi_{c}\delta\Delta_{ab}}\chi_{cef}^{\phi K}+\frac{\delta^{2}\Gamma}{\delta\Delta_{cd}\delta\Delta_{ab}}\chi_{cdef}^{\Delta K} & =-\frac{1}{2}i\hbar\left(\delta_{ae}\delta_{bf}+\delta_{af}\delta_{be}\right),
\end{align}
where note that in the last two equations we have to symmetrize the right hand sides before removing the source $K$ (since by the symmetry of $K$ only the symmetric part contributes).
These equations determine the linear response functions entirely in terms of the equilibrium properties of the theory (in particular, the second derivatives of the effective action evaluated at the equilibrium solution).
Note that the last equation can be recast as a Bethe-Salpeter equation for the $\chi^{\Delta K}$ by using (\ref{eq:2piea}) and (\ref{eq:dyson-eq}) to write the left hand side of (\ref{eq:unmodified-Delta-eom}) as
\begin{widetext}
\begin{align}
-\Delta_{ab}^{-1}+\Delta_{0ab}^{-1}\left[\varphi\right]-\Sigma_{ab}\left[\varphi,\Delta\right]	&=-\Delta_{ab}^{-1}+\left(\Delta_{0ab}^{-1}\left[\varphi\right]-\Delta_{0ab}^{-1}\left[\tilde{\varphi}\right]\right) \nonumber\\
	&+\left(\Delta_{0ab}^{-1}\left[\tilde{\varphi}\right]-\Sigma_{ab}\left[\tilde{\varphi},\tilde{\Delta}\right]\right)+\left(\Sigma_{ab}\left[\tilde{\varphi},\tilde{\Delta}\right]-\Sigma_{ab}\left[\varphi,\Delta\right]\right) \nonumber\\
	&=-\delta\Delta_{ab}^{-1}+\frac{\delta\Delta_{0ab}^{-1}}{\delta\varphi_{c}}\delta\varphi_{c}-\left(\frac{\delta\Sigma_{ab}}{\delta\varphi_{c}}\delta\varphi_{c}+\frac{\delta\Sigma_{ab}}{\delta\Delta_{cd}}\delta\Delta_{cd}\right),
\end{align}
\end{widetext}
then using the identity
\begin{equation}
\delta\Delta_{ab}^{-1}=-\tilde{\Delta}_{ac}^{-1}\delta\Delta_{cd}\tilde{\Delta}_{db}^{-1}+\mathcal{O}\left(J^{2},K^{2},JK\right),\label{eq:variation-of-inverse-delta}
\end{equation}
and the definitions of the linear response functions followed by some rearrangement to give
\begin{align}
\chi_{abef}^{\Delta K}&=-\tilde{\Delta}_{ac}\left(\delta_{ce}\delta_{df}+\delta_{cf}\delta_{de}\right)\tilde{\Delta}_{db} \nonumber\\
&-\tilde{\Delta}_{ag}\left(\frac{\delta\Delta_{0gh}^{-1}}{\delta\varphi_{c}}-\frac{\delta\Sigma_{gh}}{\delta\varphi_{c}}\right)\tilde{\Delta}_{hb}\chi_{cef}^{\phi K} \nonumber\\
&+\left(\tilde{\Delta}_{ag}\frac{\delta\Sigma_{gh}}{\delta\Delta_{cd}}\tilde{\Delta}_{hb}\right)\chi_{cdef}^{\Delta K}.
\end{align}
This is an equation which determines the four point kernel $\chi^{\Delta K}$ iteratively, i.e. a Bethe-Salpeter equation with the last quantity in braces being the Bethe-Salpeter kernel.

While this is the general formalism connecting effective actions to linear response theory, we do not need to use it because symmetry improvement constraints change the picture significantly.
In particular, we can derive the implications of the constraints for linear response theory without explicitly computing any equilibrium solutions, taking any derivatives of the effective action, performing any truncation of the self-energies, or worrying about renormalization of the Bethe-Salpeter equation.
Our results are independent of all of these details.

\section{Symmetry Improvement\label{sec:Symmetry-Improvement}}

Here we review the symmetry improvement procedure since the derivation of the constraint and the limiting procedure are affected by the presence of sources.
The idea is to force the 2PI propagator $\Delta^{-1}$ to mimic the behavior of the corresponding 1PI function $\delta^{2}\Gamma\left[\varphi\right]/\delta\varphi\delta\varphi$ under a symmetry transformation.
To derive the required WI we perform a symmetry transformation of the 1PIEA
\begin{equation}
0=\delta\Gamma=\frac{\delta\Gamma}{\delta\varphi_{a}}i\epsilon_{A}T_{ab}^{A}\varphi_{b}.
\end{equation}
(Note that our ``DeWitt'' integration convention can be maintained if we define the generators to be spacetime delta functions: $T_{ab}^{A}\left(x,y\right)\propto\delta\left(x-y\right)$.)
This is the ``master'' WI governing all 1PI correlation functions of the theory.
To find the identity governing the propagator we take a field derivative then apply the equations of motion, giving
\begin{align}
0 & =\frac{\delta^{2}\Gamma}{\delta\varphi_{c}\delta\varphi_{a}}i\epsilon_{A}T_{ab}^{A}\varphi_{b}+\frac{\delta\Gamma}{\delta\varphi_{a}}i\epsilon_{A}T_{ac}^{A}\nonumber \\
 & =\Delta_{ca}^{-1}i\epsilon_{A}T_{ab}^{A}\varphi_{b}-J_{a}i\epsilon_{A}T_{ac}^{A},
\end{align}
which holds for all rotations $\epsilon_{A}$ so we define
\begin{equation}
0=\mathcal{W}_{c}^{A}\equiv\Delta_{ca}^{-1}T_{ab}^{A}\varphi_{b}-J_{a}T_{ac}^{A}.\label{eq:WI}
\end{equation}
The symmetry improvement program then interprets the $\Delta^{-1}$ as the \emph{2PI} propagator.
One could also motivate the replacement $J_{a}\to J_{a}+K_{ab}\varphi_{b}$ based on the form of the right hand side of (\ref{eq:unmodified-vev-eom}).
We do not make this replacement.
The only affect at the linear response level would be to renormalize $J_{a}\to J_{a}+K_{aN}v$ in our results below which, since $J_{a}$ are freely chosen functions, gives no new physics.

To enforce the WI we add a Lagrange multiplier term to the 2PIEA $\Gamma\to\Gamma-\mathcal{C}$ with
\begin{equation}
\mathcal{C}=\frac{i}{2}\ell_{A}^{c}\mathcal{W}_{c}^{A},\label{eq:simple-constraint}
\end{equation}
where $\ell_{A}^{c}$ are the Lagrange multiplier fields.
In the previous paper \citep{brown_symmetry_2015} we included a transverse projector $P_{ab}^{\perp}\left(x\right)=\delta_{ab}-\varphi_{a}\left(x\right)\varphi_{b}\left(x\right)/\varphi^{2}\left(x\right)$:
\begin{equation}
\mathcal{C}'=\frac{i}{2}\ell_{A}^{c}P_{cd}^{\perp}\mathcal{W}_{d}^{A},\label{eq:projected-constraint}
\end{equation}
to ensure that only Goldstone modes are involved in the constraint.
This turns out to make no difference in equilibrium.
However, this constraint gives a different scheme beyond the equilibrium approximation.
(Note that Pilaftsis and Teresi \citep{pilaftsis_symmetry_2013} did not write the constraint in an $O\left(N\right)$ covariant form, hence both of the above are valid generalizations of their procedure.)
Using $\mathcal{C}'$ instead of $\mathcal{C}$ merely seems to shuffle around the pathologies we derive below rather than remove them.
We shall henceforth use $\mathcal{C}$ since dropping the projector greatly simplifies the following algebra.
See Appendix \ref{sec:The-Alternate-Constraint} for discussion of the scheme using $\mathcal{C}'$.

The equations of motion following from the symmetry improved effective action are
\begin{widetext}
\begin{align}
\mathcal{W}_{c}^{A} & =0,\\
\frac{\delta\Gamma}{\delta\varphi_{d}\left(z\right)} & =\frac{i}{2}\int_{x}\ell_{A}^{c}\left(x\right)\Delta_{ca}^{-1}\left(x,z\right)T_{ad}^{A}-J_{d}\left(z\right)-\int_{w}K_{de}\left(z,w\right)\varphi_{e}\left(w\right),\label{eq:si-eom-vev}\\
\frac{\delta\Gamma}{\delta\Delta_{de}\left(z,w\right)} & =\frac{i}{2}\int_{x}\ell_{A}^{c}\left(x\right)\int_{y}\frac{\delta\Delta_{ca}^{-1}\left(x,y\right)}{\delta\Delta_{de}\left(z,w\right)}T_{ab}^{A}\varphi_{b}\left(y\right)-\frac{1}{2}i\hbar K_{de}\left(z,w\right),
\end{align}
where the last equation simplifies to
\begin{equation}
\frac{\delta\Gamma}{\delta\Delta_{de}\left(z,w\right)}=-\frac{i}{2}\int_{x}\ell_{A}^{c}\left(x\right)\Delta_{cd}^{-1}\left(x,z\right)\int_{y}\Delta_{ea}^{-1}\left(w,y\right)T_{ab}^{A}\varphi_{b}\left(y\right)-\frac{1}{2}i\hbar K_{de}\left(z,w\right),\label{eq:si-eom-delta}
\end{equation}
on using the identity $\delta\Delta_{ca}^{-1}/\delta\Delta_{de}=-\Delta_{cd}^{-1}\Delta_{ea}^{-1}$.
\end{widetext}

We now recall what happens to the right hand sides of these equations in equilibrium with $J=K=0$ (this is reviewing \citep{pilaftsis_symmetry_2013,brown_symmetry_2015}).
The only non-trivial WIs are $\mathcal{W}_{c}^{gN}=-ivm_{G}^{2}P_{cg}^{\perp}$, so that the constraint enforces $vm_{G}^{2}=0$, i.e. the Goldstone mass vanishes if $v\neq0$ as expected.
Using homogeneity $\ell_{A}^{c}\left(x\right)=\ell_{A}^{c}$, and equations (\ref{eq:eq-ssb-ansatz-vev})-(\ref{eq:eq-ssb-ansatz-delta}) the other equations of motion become
\begin{align}
\frac{\partial\Gamma/VT}{\partial v} & =\ell_{cN}^{c}m_{G}^{2},\\
\frac{\delta\Gamma}{\delta\Delta_{G}\left(z,w\right)} & =\ell_{cN}^{c}vm_{G}^{4},\\
\frac{\delta\Gamma}{\delta\Delta_{H}\left(z,w\right)} & =0,
\end{align}
where $VT$ is the volume of spacetime.
Now, since $m_{G}^{2}\to0$ the right hand sides vanish unless $\ell_{cN}^{c}\to\infty$.
We can consistently set
\begin{equation}
\ell_{bN}^{a}=-\ell_{Nb}^{a}=P_{ab}^{\perp}\left(\frac{1}{N-1}\ell_{cN}^{c}\right),\label{eq:lagrange-multipliers}
\end{equation}
and all other components of $\ell_{A}^{c}$ to zero.
Thus the constraint is a singular one.
We regulate the divergence by setting $vm_{G}^{2}=\eta m^{3}$ and take the limit $\eta\to0$ such that $\eta\ell_{cN}^{c}/v=\ell_{0}$ is a constant.
This gives
\begin{align}
\frac{\partial\Gamma/VT}{\partial v} & =\ell_{0}m^{3},\\
\frac{\delta\Gamma}{\delta\Delta_{G}\left(z,w\right)} & =0.
\end{align}
Thus the propagator equations of motion are unmodified and the vev equation is modified by the presence of a homogeneous force that acts to push $v$ away from the minimum of the effective potential to the point where $m_{G}^{2}=0$.

\section{Implications of Constraints for Linear Response\label{sec:Implications-of-Constraints}}

In the linear response approximation $J_{a}$ and $K_{ab}$ no longer vanish, and we must solve equations (\ref{eq:WI}), (\ref{eq:si-eom-vev}) and (\ref{eq:si-eom-delta}) to first order in the sources without any assumption of homogeneity.
To do this we expand all quantities $\varphi\to\tilde{\varphi}+\delta\varphi$, $\Delta\to\tilde{\Delta}+\delta\Delta$ and $\ell\to\tilde{\ell}+\delta\ell$ and match terms order by order, considering the $\delta\varphi$ etc. as first order.
Working first on (\ref{eq:WI}) gives the pair of equations:
\begin{align}
0 & =\tilde{\Delta}_{ca}^{-1}T_{ab}^{A}\tilde{\varphi}_{b},\label{eq:WI-zeroth-order}\\
0 & =\delta\Delta_{ca}^{-1}T_{ab}^{A}\tilde{\varphi}_{b}+\tilde{\Delta}_{ca}^{-1}T_{ab}^{A}\delta\varphi_{b}-J_{a}T_{ac}^{A}.\label{eq:WI-first-order}
\end{align}
The first equation is simply the equilibrium constraint as expected since the analysis of the last section holds with all quantities decorated with tildes.
The second equation is new.
Using the identity (\ref{eq:variation-of-inverse-delta}) gives to first order
\begin{equation}
0=-\tilde{\Delta}_{cd}^{-1}\delta\Delta_{de}\left(\tilde{\Delta}_{ea}^{-1}T_{ab}^{A}\tilde{\varphi}_{b}\right)+\tilde{\Delta}_{ca}^{-1}T_{ab}^{A}\delta\varphi_{b}-J_{a}T_{ac}^{A}.
\end{equation}
The first term vanishes by virtue of (\ref{eq:WI-zeroth-order}), thus
\begin{equation}
\tilde{\Delta}_{ca}^{-1}T_{ab}^{A}\delta\varphi_{b}=J_{a}T_{ac}^{A}.\label{eq:first-order-constraint-simple-scheme}
\end{equation}
The nature of this equation is remarkable.
The \emph{constraint} yields a \emph{wave equation} for the fluctuations.
Working out the component equations gives:
\begin{align}
\tilde{\Delta}_{G}^{-1}\delta\varphi_{N} & =-J_{N}, & A=\left(g,N\right),\ c\neq N,\label{eq:phiN-eom-massless}\\
\tilde{\Delta}_{H}^{-1}\delta\varphi_{g} & =-J_{g}, & A=\left(g,N\right),\ c=N,\label{eq:phiG-eom-w-Higgs-propagator}\\
\tilde{\Delta}_{G}^{-1}\delta\varphi_{g} & =-J_{g}, & A=\left(g,g'\right),\ g,g'\neq N,\ c=g', c\neq g.\label{eq:phiG-eom-w-Goldstone-propagator}
\end{align}
The desired linear response functions are then just the Green functions for these equations, i.e. $\chi_{NN}^{\phi J}\left(x,y\right)=-\tilde{\Delta}_{G}\left(x,y\right)$ etc.

There are two major flaws with these equations.
The first is a physical misprediction: $\delta\varphi_{N}$ propagates masslessly due to the equilibrium Goldstone propagator even though the Higgs propagator has a non-zero mass.
The second is that $\delta\varphi_{g}$ is overdetermined, i.e. its initial value problem is ill posed.
In the Fourier domain equations (\ref{eq:phiG-eom-w-Higgs-propagator}) and (\ref{eq:phiG-eom-w-Goldstone-propagator}) read
\begin{eqnarray}
\left(p^{2}-m_{H}^{2}-\tilde{\Sigma}_{H}\left(p\right)\right)\delta\varphi_{g} & = & -J_{g},\\
\left(p^{2}-\tilde{\Sigma}_{G}\left(p\right)\right)\delta\varphi_{g} & = & -J_{g},
\end{eqnarray}
which obviously only allows solutions if there are modes satisfying
$m_{H}^{2}+\tilde{\Sigma}_{H}\left(p\right)=\tilde{\Sigma}_{G}\left(p\right)$
and if $J_{g}$ is only supported on these modes.
If any other perturbation is applied or a solution of $m_{H}^{2}+\tilde{\Sigma}_{H}\left(p\right)=\tilde{\Sigma}_{G}\left(p\right)$ does not exist the system is inconsistent.
This is clearly not the expected behavior physically.
One can relax (\ref{eq:phiG-eom-w-Higgs-propagator}) by using the projected constraint (\ref{eq:projected-constraint}) instead of (\ref{eq:simple-constraint}), however this leads to further difficulties as discussed in Appendix \ref{sec:The-Alternate-Constraint}.

There is further danger lurking in the right hand sides of equations (\ref{eq:si-eom-vev}) and (\ref{eq:si-eom-delta}).
Since the constraint procedure involves a limit $\tilde{\ell}\to\infty$ there is a danger that terms on the right hand sides can diverge.
Consider the expansion of the right hand side of (\ref{eq:si-eom-vev}) to first order:
\begin{multline}
\frac{i}{2}\int_{x}\delta\ell_{A}^{c}\left(x\right)\tilde{\Delta}_{ca}^{-1}\left(x,z\right)T_{ad}^{A}+\frac{i}{2}\int_{x}\tilde{\ell}_{A}^{c}\left(x\right)\delta\Delta_{ca}^{-1}\left(x,z\right)T_{ad}^{A}\nonumber \\
-J_{d}\left(z\right)-\int_{w}K_{de}\left(z,w\right)\tilde{\varphi}_{e}\left(w\right).
\end{multline}
Using (\ref{eq:lagrange-multipliers}) the term proportional to $\tilde{\ell}$ can be written
\begin{equation}
-\frac{1}{N-1}\tilde{\ell}_{eN}^{e}\left(P_{cg}^{\perp}\int_{x}\delta\Delta_{cg}^{-1}\left(x,z\right)\delta_{Nd}-P_{cd}^{\perp}\int_{x}\delta\Delta_{cN}^{-1}\left(x,z\right)\right),
\end{equation}
which, for the $\tilde{\ell}_{eN}^{e}\to\infty$ limit to exist, requires that the term in braces vanishes, i.e.
\begin{equation}
0=\delta_{Nd}\int_{x}P_{cg}^{\perp}\delta\Delta_{cg}^{-1}\left(x,z\right)-P_{cd}^{\perp}\int_{x}\delta\Delta_{cN}^{-1}\left(x,z\right).\label{eq:secondary-constraint-phi-eom}
\end{equation}
A similar analysis for (\ref{eq:si-eom-delta}) gives
\begin{align}
0 & =vm_{G}^{2}\int_{x}\left(P_{ea}^{\perp}\delta\Delta_{ad}^{-1}\left(x,z\right)+\delta\Delta_{ea}^{-1}\left(w,x\right)P_{ad}^{\perp}\right)\nonumber \\
 & -P_{db}^{\perp}\delta_{eN}m_{G}^{2}\int_{y}\tilde{\Delta}_{H}^{-1}\left(w,y\right)\delta\varphi_{b}\left(y\right) \nonumber \\
 & +P_{de}^{\perp}m_{G}^{2}\int_{y}\tilde{\Delta}_{G}^{-1}\left(w,y\right)\delta\varphi_{N}\left(y\right).\label{eq:secondary-constraint-delta-eom}
\end{align}
We call (\ref{eq:secondary-constraint-phi-eom}) and (\ref{eq:secondary-constraint-delta-eom}) the \emph{secondary} constraints of the scheme and, by contrast, equation (\ref{eq:WI-first-order}) the \emph{primary} constraint \footnote{These should not to be confused with the terminology from Dirac's constrained quantization method.}.
The secondary constraints must be enforced so that no divergences appear in the $\tilde{\ell}\to\infty$ limit of the equations of motion.

Note that one can take $m_{G}^{2}\to0$ without any problems in (\ref{eq:secondary-constraint-delta-eom}), so that constraint is satisfied identically. Similarly, using (\ref{eq:variation-of-inverse-delta}), (\ref{eq:secondary-constraint-phi-eom}) becomes
\begin{align}
0&=m_{G}^{2}\int_{yw}\left(\delta_{dN}P_{cg}^{\perp}\delta\Delta_{cg}\left(y,w\right)\tilde{\Delta}_{G}^{-1}\left(w,z\right) \right. \nonumber \\
 &\left.-P_{de}^{\perp}\delta\Delta_{eN}\left(y,w\right)\tilde{\Delta}_{H}^{-1}\left(w,z\right)\right),
\end{align}
which is also automatically satisfied in the $m_{G}^{2}\to0$ limit.
The fact that the secondary constraints automatically vanish is a consequence of using the unprojected constraint (\ref{eq:simple-constraint}).
Had we used (\ref{eq:projected-constraint}) instead, the troublesome equation of motion (\ref{eq:phiG-eom-w-Higgs-propagator}) would be gone, but the secondary constraints become non-trivial and lead again to pathologies (see Appendix \ref{sec:The-Alternate-Constraint}).

\section{Discussion\label{sec:Discussion}}

We have shown that the imposition of symmetry improvement constraints is incompatible with the linear response approximation.
Since the original symmetry improvement scheme of Pilaftsis and Teresi was not formulated $O\left(N\right)$-covariantly there are actually two natural generalizations: a scheme we have used previously \citep{brown_symmetry_2015} based on (\ref{eq:projected-constraint}) and a new one based on the simpler constraint (\ref{eq:simple-constraint}).
Both schemes are equivalent in equilibrium and both lead, in different ways, to pathologies in the linear response approximation.

There are two types of pathology appearing in our results.
The first is that the Higgs field fluctuations propagate masslessly.
The second and more serious pathology is that the Goldstone field fluctuations are over-determined, and this over-determination happens in both constraint schemes we consider (though in different ways).
There is no simple modification of the constraint which could possibly fix these problems since we can understand them as a consequence of treating $\varphi$ and $\Delta$ as independent variables in the 2PIEA and the inability of symmetry improved 2PIEA to enforce also the three point \emph{vertex} WI.

To see these aspects of the problem consider the first order Ward identity for the 1PIEA, which is the same in form as (\ref{eq:WI-first-order}),
\begin{equation}
0=\delta\Delta_{ca}^{-1}T_{ab}^{A}\tilde{\varphi}_{b}+\tilde{\Delta}_{ca}^{-1}T_{ab}^{A}\delta\varphi_{b}-J_{a}T_{ac}^{A},
\end{equation}
only now $\Delta$ is not independent and $\delta\Delta^{-1}$ is to be understood as arising purely from the variation $\delta\varphi$.
To lowest order in $\hbar$:
\begin{align}
\delta\Delta_{ca}^{-1}& =\delta\Delta_{0ca}^{-1} \nonumber \\
 & =\left(-\frac{\lambda}{3}\tilde{\varphi}_{d}\delta\varphi^{d}\delta_{ac}-\frac{\lambda}{3}\delta\varphi_{a}\tilde{\varphi}_{c}-\frac{\lambda}{3}\tilde{\varphi}_{a}\delta\varphi_{c}\right)\delta\left(x-y\right)\nonumber \\
 & =-\frac{\lambda}{3}v\left(\delta\varphi_{N}\delta_{ac}+\delta\varphi_{a}\delta_{cN}+\delta_{aN}\delta\varphi_{c}\right)\delta\left(x-y\right).
\end{align}
Substituting this into the WI gives
\begin{align}
0&=-\frac{\lambda v^{2}}{3}\left(\delta\varphi_{N}\left(x\right)\delta_{ac}+\delta\varphi_{a}\left(x\right)\delta_{cN}\right)T_{aN}^{A} \nonumber \\
 &+\int_{y}\tilde{\Delta}_{ca}^{-1}\left(x,y\right)T_{ab}^{A}\delta\varphi_{b}\left(y\right)-J_{a}T_{ac}^{A},
\end{align}
and working out the components we get the set of equations
\begin{align}
\int_{y}\left[\tilde{\Delta}_{G}^{-1}\left(x,y\right)-\frac{\lambda v^{2}}{3}\delta\left(x-y\right)\right]\delta\varphi_{N}\left(y\right) & =-J_{N},\\
\int_{y}\left[\tilde{\Delta}_{H}^{-1}\left(x,y\right)+\frac{\lambda v^{2}}{3}\delta\left(x-y\right)\right]\delta\varphi_{g}\left(y\right) & =-J_{g},\\
\int_{y}\tilde{\Delta}_{G}^{-1}\left(x,y\right)\delta\varphi_{g}\left(y\right) & =-J_{g},
\end{align}
which is to be contrasted with (\ref{eq:phiN-eom-massless})-(\ref{eq:phiG-eom-w-Goldstone-propagator}).
This system is consistent if $m_{H}^{2}=m_{G}^{2}+\lambda v^{2}/3+\mathcal{O}\left(\hbar\right)$ which, of course, is true.
Thus, for the 1PIEA the WI propagates Higgs and Goldstone fluctuations with the correct masses and source terms.
This is expected because the WI was constructed to be satisfied by the 1PI correlation functions.
There is no longer any reason for this to work if $\delta\Delta$ is independent of $\delta\varphi$.

We can extend this analysis by considering (\ref{eq:WI-first-order}) again.
From the definition in terms of the 1PIEA, $\Delta^{-1} = \delta^2 \Gamma^{(1)} / \delta \varphi \delta \varphi$, we can write
\begin{equation}
\delta\Delta_{ca}^{-1}=\frac{\delta^{3}\Gamma^{\left(1\right)}}{\delta\varphi_{d}\delta\varphi_{c}\delta\varphi_{a}}\delta\varphi_{d}+\mathcal{O}\left(\delta\varphi^{2}\right),
\end{equation}
and note that $\delta^{3}\Gamma^{\left(1\right)}/\delta\varphi_{d}\delta\varphi_{c}\delta\varphi_{a}=V_{dca}$ is the three point vertex function.
These relations no longer hold identically for the 2PI correlation functions, but they do hold numerically for the \emph{exact} solutions of the untruncated 2PI equations of motion.
We can now extend this to 2PIEA by writing
\begin{equation}
\delta\Delta_{ca}^{-1}=V_{dca}\delta\varphi_{d}+\mathcal{K}_{ca},
\end{equation}
 where $\mathcal{K}_{ca}$ encapsulates the additional variations of $\Delta$ in the truncated 2PIEA formalism.

There is a WI for the three point vertex function (c.f. equation (3.6) of \cite{brown_symmetry_2015}),
\begin{equation}
\mathcal{W}_{dc}^{A}=V_{dca}T_{ab}^{A}\varphi_{b}+\Delta_{ca}^{-1}T_{ad}^{A}+\Delta_{da}^{-1}T_{ac}^{A},
\end{equation}
which is unaffected by the presence of sources.
As with (\ref{eq:WI}), $\mathcal{W}_{dc}^{A}=0$ in the exact theory, but not automatically in truncations.
Combining these relations with  (\ref{eq:WI-first-order}) gives:
\begin{align}
0 &=\delta\varphi_{d}\left(\mathcal{W}_{dc}^{A}-\tilde{\Delta}_{ca}^{-1}T_{ad}^{A}-\tilde{\Delta}_{da}^{-1}T_{ac}^{A}\right)+\mathcal{K}_{ca}T_{ab}^{A}\tilde{\varphi}_{b} \nonumber \\
 &+\tilde{\Delta}_{ca}^{-1}T_{ab}^{A}\delta\varphi_{b}-J_{a}T_{ac}^{A} \nonumber \\
 &=\delta\varphi_{d}\left(\mathcal{W}_{dc}^{A}-\tilde{\Delta}_{da}^{-1}T_{ac}^{A}\right)+\mathcal{K}_{ca}T_{ab}^{A}\tilde{\varphi}_{b}-J_{a}T_{ac}^{A},
\end{align}
which can be rearranged as
\begin{equation}
\left(\delta\varphi_{d}\tilde{\Delta}_{da}^{-1}+J_{a}\right)T_{ac}^{A}=\delta\varphi_{d}\mathcal{W}_{dc}^{A}+\mathcal{K}_{ca}T_{ab}^{A}\tilde{\varphi}_{b}. \label{eq:fluctuation-eom-vertex-WI}
\end{equation} 

If the right hand side vanishes one has (using the symmetry of $\tilde{\Delta}^{-1}$)
\begin{equation}
\tilde{\Delta}_{ad}^{-1}\delta\varphi_{d}=-J_{a},
\end{equation}
which is the correct equation of motion for fluctuations.
The failure of these equations to be satisfied is measured by the terms on the right hand side of (\ref{eq:fluctuation-eom-vertex-WI}), which have two distinct meanings.
The first, $\delta\varphi_{d}\mathcal{W}_{dc}^{A}$, measures the failure of the \emph{vertex} WI to be satisfied by 2PI approximations, while the second term, $\mathcal{K}_{ca}T_{ab}^{A}\tilde{\varphi}_{b}$, measures the failure of the variations in the 2PI functions $\varphi$ and $\Delta$ to be linked according to the 1PI relations.

One could try to eliminate the second error term by constraining the variation $\delta\Delta$ to be related to $\delta\varphi$ in a appropriate way.
This would no longer be working strictly within the 2PIEA formalism.
Rather, it would define a hybrid 2PI-1PI scheme where one computes the equilibrium properties using the symmetry improved 2PIEA, then defines a resummed 1PIEA by eliminating $\Delta$ from $\Gamma$ in an appropriate way.
This is similar to the usual resummed 1PIEA method, only now symmetry improvement is applied self-consistently at the 2PI level.

The first error term is more formidable.
It comes down to the failure of the master 1PI WI in $n$PIEA truncations.
The master WI encodes relations between \emph{all} correlation functions in the exact theory.
However, a symmetry improved $n$PIEA scheme only has the power to enforce constraints between the lowest $n$ of them.
Violations of WIs involving higher order correlation functions are inevitable in any scheme with fixed finite $n$.
This is not a major issue in equilibrium because one can solve for the $n$-point correlation functions in a self-consistently complete $n$-loop truncation, and so long as one does not care about the behavior of higher order correlation functions their problems remain invisible.
However, once external sources are turned on and the system departs from equilibrium, variations of the correlation functions appear and these can be related to higher order correlation functions.
Violations of the higher order Ward identities then feedback into the equations of motion for the fluctuations, leading to an inconsistent system overall.
This leads to our general conclusion: \emph{all symmetry improved $n$PIEA schemes are incompatible with the linear response approximation}.

It is interesting that the same number of non-trivial constraints is found in both schemes, despite the attempt to reduce their number in the projected scheme.
This is because the projection operator depends on the fields $\varphi$, so it has non-trivial derivatives.
This will be the case for any $O\left(N\right)$-covariant projection scheme.
This motivates a study of whether it is possible to \emph{non}-covariantly project out the troublesome constraints, i.e. explicitly break the symmetry in order to save it.

It is worth mentioning the logical possibility that the true behavior of the solutions to the truncated SI-2PIEA equations of motion, when solutions exist, is non-analytic as a function of the sources in the neighborhood of equilibrium.
This would invalidate any attempted Taylor series expansion in the sources and so we should expect problems at the linear response level.
We do not know at this time how to further analyze this possibility.

It would also be interesting to examine whether alternatives to symmetry improvement, such as the method for enforcing constraints using external sources \citep{garbrecht_constraining_2016}, can be extended to non-equilibrium situations.
We plan in future work to investigate a scheme with \emph{softy} imposed symmetry improvement as a potential workaround for the issues found here and in \citep{marko_loss_2016}.

\appendix

\section{The Alternate Constraint Scheme\label{sec:The-Alternate-Constraint}}

Here we consider using the constraint (\ref{eq:projected-constraint}) instead of (\ref{eq:simple-constraint}) for the symmetry improvement.
The projection operator in (\ref{eq:projected-constraint}) forces only Goldstone modes to be involved in the constraint.
Note that the constraints which have been projected out are valid WIs which are obeyed by the 1PIEA.
This alternate scheme consists of picking a subset of the WIs to enforce, chosen in the only $O\left(N\right)$-covariant way available.
The new constraint is
\begin{equation}
0=P_{cd}^{\perp}\left(\Delta_{da}^{-1}T_{ab}^{A}\varphi_{b}+T_{da}^{A}J_{a}\right).
\end{equation}
At lowest order this becomes
\begin{equation}
0=-vm_{G}^{2}T_{cN}^{A},
\end{equation}
which is the same as before.
However, at first order there are new terms due to the variation of $P^{\perp}$.
For reference we give the first and second derivatives of $P^{\perp}$ evaluated at $\tilde{\varphi}$:
\begin{widetext}
\begin{align}
\frac{\delta P_{dc}^{\perp}\left(x\right)}{\delta\varphi_{e}\left(z\right)}\left[\tilde{\varphi}\right] & =-\frac{1}{v}\left(\delta_{de}\delta_{cN}+\delta_{dN}\delta_{ce}-2\delta_{dN}\delta_{cN}\delta_{eN}\right)\delta\left(x-z\right),\\
\frac{\delta^{2}P_{dc}^{\perp}\left(x\right)}{\delta\varphi_{f}\left(w\right)\delta\varphi_{e}\left(z\right)}\left[\tilde{\varphi}\right] & =\frac{1}{v^{2}}F_{dcfe}\delta\left(z-w\right)\delta\left(x-z\right),
\end{align}
where
\begin{align}
F_{dcfe} & =2\delta_{cN}\delta_{dN}\tilde{P}_{ef}^{\perp}-\left(\tilde{P}_{ce}^{\perp}\tilde{P}_{df}^{\perp}+\tilde{P}_{de}^{\perp}\tilde{P}_{cf}^{\perp}\right)\nonumber \\
 & +\left(\tilde{P}_{ce}^{\perp}\delta_{dN}\delta_{fN}+\tilde{P}_{de}^{\perp}\delta_{cN}\delta_{fN}+\tilde{P}_{cf}^{\perp}\delta_{dN}\delta_{eN}+\tilde{P}_{df}^{\perp}\delta_{cN}\delta_{eN}\right),
\end{align}
which is symmetric in $cd$ and in $ef$.
In the linear response approximation we need both
\begin{align}
\delta P_{cd}^{\perp}\left(x\right) & \to\int_{z}\frac{\delta P_{dc}^{\perp}\left(x\right)}{\delta\varphi_{e}\left(z\right)}\left[\tilde{\varphi}\right]\delta\varphi_{e}\left(z\right)=-\frac{1}{v}\left(\delta_{cN}\delta\varphi_{d}\left(x\right)+\delta_{dN}\delta\varphi_{c}\left(x\right)-2\delta_{dN}\delta_{cN}\delta\varphi_{N}\left(x\right)\right),\\
\delta\left[\frac{\delta P_{cd}^{\perp}\left(x\right)}{\delta\varphi_{e}\left(z\right)}\right] & \to\int_{w}\frac{\delta^{2}P_{dc}^{\perp}\left(x\right)}{\delta\varphi_{f}\left(w\right)\delta\varphi_{e}\left(z\right)}\left[\tilde{\varphi}\right]\delta\varphi_{f}\left(w\right)=\frac{1}{v^{2}}F_{dcfe}\delta\left(x-z\right)\delta\varphi_{f}\left(x\right).
\end{align}
\end{widetext}

The first order constraint is
\begin{align}
0&=P_{cd}^{\perp}\left(\delta\Delta_{da}^{-1}T_{ab}^{A}\tilde{\varphi}_{b}+\tilde{\Delta}_{da}^{-1}T_{ab}^{A}\delta\varphi_{b}+T_{da}^{A}J_{a}\right) \nonumber \\
 &+\delta P_{cd}^{\perp}\left(\tilde{\Delta}_{da}^{-1}T_{ab}^{A}\tilde{\varphi}_{b}\right).
\end{align}
The second term vanishes due to (\ref{eq:WI-zeroth-order}) and the first term is just (\ref{eq:first-order-constraint-simple-scheme}) with only the $c\neq N$ equations picked out.
This gives the set of wave equations:
\begin{align}
\tilde{\Delta}_{G}^{-1}\delta\varphi_{N} & =-J_{N},\label{eq:phiN-eom-massless-1}\\
\tilde{\Delta}_{G}^{-1}\delta\varphi_{g} & =-J_{g},\label{eq:phiG-eom-w-Goldstone-propagator-1}
\end{align}
which would be the end of the story if not for the secondary constraints.
However, do note that the Higgs field still propagates masslessly.

In order to find the secondary constraints we write the equations of motion
\begin{widetext}
\begin{align}
\frac{\delta\Gamma}{\delta\varphi_{d}\left(z\right)} & =\frac{i}{2}\int_{x}\ell_{A}^{f}\left(x\right)\frac{\delta P_{fc}^{\perp}\left(x\right)}{\delta\varphi_{d}\left(z\right)}\left(\int_{y}\Delta_{ca}^{-1}\left(x,y\right)T_{ab}^{A}\varphi_{b}\left(y\right)+T_{ca}^{A}J_{a}\left(x\right)\right)\nonumber \\
 & +\frac{i}{2}\int_{x}\ell_{A}^{f}\left(x\right)P_{fc}^{\perp}\left(x\right)\Delta_{ca}^{-1}\left(x,z\right)T_{ad}^{A}-J_{d}\left(z\right)-\int_{w}K_{de}\left(z,w\right)\varphi_{e}\left(w\right),\label{eq:si-eom-phi-projected-scheme}\\
\frac{\delta\Gamma}{\delta\Delta_{de}\left(z,w\right)} & =-\frac{i}{2}\int_{x}\ell_{A}^{f}\left(x\right)P_{fc}^{\perp}\left(x\right)\Delta_{cd}^{-1}\left(x,z\right)\int_{y}\Delta_{ea}^{-1}\left(w,y\right)T_{ab}^{A}\varphi_{b}\left(y\right)-\frac{1}{2}i\hbar K_{de}\left(z,w\right).\label{eq:si-eom-delta-projected-scheme}
\end{align}
The secondary constraints are that the variation of the terms multiplying $\ell_{A}^{f}$ vanish, since if they did not divergences would arise as $\tilde{\ell}\to\infty$.
We start work on the right hand side of (\ref{eq:si-eom-delta-projected-scheme}) by demanding
\begin{align}
0 & =-\frac{i}{2}\int_{x}\tilde{\ell}_{A}^{f}\delta\left[P_{fc}^{\perp}\left(x\right)\Delta_{cd}^{-1}\left(x,z\right)\int_{y}\Delta_{ea}^{-1}\left(w,y\right)T_{ab}^{A}\varphi_{b}\left(y\right)\right]\nonumber \\
 & =-i\frac{1}{N-1}\tilde{\ell}_{hN}^{h}\tilde{P}_{gf}^{\perp}\int_{x}\delta\left[P_{fc}^{\perp}\left(x\right)\Delta_{cd}^{-1}\left(x,z\right)\int_{y}\Delta_{ea}^{-1}\left(w,y\right)T_{ab}^{gN}\varphi_{b}\left(y\right)\right],
\end{align}
which gives the constraint
\begin{align}
0 & =-i\tilde{P}_{gf}^{\perp}\int_{x}\delta\left[P_{fc}^{\perp}\left(x\right)\Delta_{cd}^{-1}\left(x,z\right)\int_{y}\Delta_{ea}^{-1}\left(w,y\right)T_{ab}^{gN}\varphi_{b}\left(y\right)\right]\nonumber \\
 & =\tilde{P}_{ef}^{\perp}\int_{x}\delta P_{fc}^{\perp}\left(x\right)\tilde{\Delta}_{cd}^{-1}\left(x,z\right)\int_{y}\tilde{\Delta}_{G}^{-1}\left(w,y\right)v+\tilde{P}_{ec}^{\perp}\int_{x}\delta\Delta_{cd}^{-1}\left(x,z\right)\int_{y}\tilde{\Delta}_{G}^{-1}\left(w,y\right)v\nonumber \\
 & +\tilde{P}_{ad}^{\perp}\int_{x}\tilde{\Delta}_{G}^{-1}\left(x,z\right)\int_{y}\delta\Delta_{ea}^{-1}\left(w,y\right)v-\delta_{eN}\tilde{P}_{gd}^{\perp}\int_{x}\tilde{\Delta}_{G}^{-1}\left(x,z\right)\int_{y}\tilde{\Delta}_{H}^{-1}\left(w,y\right)\delta\varphi_{g}\left(y\right)\nonumber \\
 & +\tilde{P}_{ed}^{\perp}\int_{x}\tilde{\Delta}_{G}^{-1}\left(x,z\right)\int_{y}\tilde{\Delta}_{G}^{-1}\left(w,y\right)\delta\varphi_{N}\left(y\right),
\end{align}
we find again that every term is proportional to $m_{G}^{2}\to0$ so the constraint is satisfied automatically.

Now working on the right hand side of (\ref{eq:si-eom-phi-projected-scheme}) gives the secondary constraint
\begin{align}
0 & =i\tilde{P}_{gf}^{\perp}\int_{x}\delta\left[\frac{\delta P_{fc}^{\perp}\left(x\right)}{\delta\varphi_{d}\left(z\right)}\left(\int_{y}\Delta_{ca}^{-1}\left(x,y\right)T_{ab}^{gN}\varphi_{b}\left(y\right)+T_{ca}^{gN}J_{a}\left(x\right)\right)+P_{fc}^{\perp}\left(x\right)\Delta_{ca}^{-1}\left(x,z\right)T_{ad}^{gN}\right]\nonumber \\
 & =i\tilde{P}_{gf}^{\perp}\int_{x}\delta\left[\frac{\delta P_{fc}^{\perp}\left(x\right)}{\delta\varphi_{d}\left(z\right)}\right]\int_{y}\tilde{\Delta}_{ca}^{-1}\left(x,y\right)T_{ab}^{gN}\tilde{\varphi}_{b}\left(y\right)
  +i\tilde{P}_{gf}^{\perp}\int_{x}\frac{\delta P_{fc}^{\perp}\left(x\right)}{\delta\varphi_{d}\left(z\right)}\left[\tilde{\varphi}\right]\int_{y}\delta\Delta_{ca}^{-1}\left(x,y\right)T_{ab}^{gN}\tilde{\varphi}_{b}\left(y\right)\nonumber \\
 & +i\tilde{P}_{gf}^{\perp}\int_{x}\frac{\delta P_{fc}^{\perp}\left(x\right)}{\delta\varphi_{d}\left(z\right)}\left[\tilde{\varphi}\right]\int_{y}\tilde{\Delta}_{ca}^{-1}\left(x,y\right)T_{ab}^{gN}\delta\varphi_{b}\left(y\right)
  +i\tilde{P}_{gf}^{\perp}\int_{x}\frac{\delta P_{fc}^{\perp}\left(x\right)}{\delta\varphi_{d}\left(z\right)}\left[\tilde{\varphi}\right]T_{ca}^{gN}J_{a}\left(x\right)\nonumber \\
 & +i\tilde{P}_{gf}^{\perp}\int_{x}\delta P_{fc}^{\perp}\left(x\right)\tilde{\Delta}_{ca}^{-1}\left(x,z\right)T_{ad}^{gN}
  +i\tilde{P}_{gf}^{\perp}\int_{x}\tilde{P}_{fc}^{\perp}\left(x\right)\delta\Delta_{ca}^{-1}\left(x,z\right)T_{ad}^{gN}.
\end{align}
\end{widetext}
(Note that the $J_{a}$ term in the third line is present because in the first line the $\delta\left[\cdots\right]$ truly means ``linear piece of $\left[\cdots\right]$'', not ``variation of $\left[\cdots\right]$.'')
Plugging in the expressions for $\delta P^{\perp}$, $\delta P^{\perp}/\delta\varphi$ and $\delta\left[\delta P^{\perp}/\delta\varphi\right]$ and simplifying gives
\begin{align}
0 & =\frac{1}{v}F_{fchd}\delta\varphi_{h}\left(z\right)m_{G}^{2}P_{fc}^{\perp}+\int_{y}\delta\Delta_{Na}^{-1}\left(z,y\right)P_{ad}^{\perp}\nonumber \\
 & -\frac{1}{v}\int_{y}\tilde{\Delta}_{H}^{-1}\left(z,y\right)P_{db}^{\perp}\delta\varphi_{b}\left(y\right)-\frac{1}{v}P_{da}^{\perp}J_{a}\left(z\right)\nonumber \\
 & -\frac{1}{v}\int_{x}\delta\varphi_{g}\left(x\right)\tilde{\Delta}_{H}^{-1}\left(x,z\right)P_{gd}^{\perp}+i\int_{x}\delta\Delta_{ga}^{-1}\left(x,z\right)T_{ad}^{gN}.
\end{align}
The first term vanishes as $m_{G}^{2}\to0$.
The second term also vanishes as
\begin{align}
\int_{y}\delta\Delta_{Na}^{-1}\left(z,y\right)P_{ad}^{\perp} & =-\int_{wvy}\tilde{\Delta}_{Nb}^{-1}\left(z,w\right)\delta\Delta_{bc}\left(w,v\right)\tilde{\Delta}_{ca}^{-1}\left(v,y\right)P_{ad}^{\perp}\nonumber \\
 & =-\int_{wv}\tilde{\Delta}_{Nb}^{-1}\left(z,w\right)\delta\Delta_{bc}\left(w,v\right)P_{cd}^{\perp} \nonumber \\
 & \times \int_{y}\tilde{\Delta}_{G}^{-1}\left(v,y\right)\nonumber \\
 & =\int_{wv}\tilde{\Delta}_{Nb}^{-1}\left(z,w\right)\delta\Delta_{bc}\left(w,v\right)P_{cd}^{\perp}m_{G}^{2}.
\end{align}

The remaining terms become, in the case $d=N$:
\begin{align}
0 & =\int_{x}\delta\Delta_{ga}^{-1}\left(x,z\right)P_{ga}^{\perp}\nonumber \\
 & =-\int_{xyw}\tilde{\Delta}_{gb}^{-1}\left(x,y\right)\delta\Delta_{bc}\left(y,w\right)\tilde{\Delta}_{ca}^{-1}\left(w,z\right)P_{ga}^{\perp}\nonumber \\
 & =m_{G}^{2}\int_{yw}\delta\Delta_{bc}\left(y,w\right)\tilde{\Delta}_{G}^{-1}\left(w,z\right)P_{bc}^{\perp},
\end{align}
which is satisfied identically.
The remaining constraint is for $d\neq N$:
\begin{align}
0 & =-\frac{1}{v}\int_{y}\tilde{\Delta}_{H}^{-1}\left(z,y\right)\delta\varphi_{d}\left(y\right)-\frac{1}{v}J_{d}\left(z\right)\nonumber \\
 & -\frac{1}{v}\int_{x}\delta\varphi_{d}\left(x\right)\tilde{\Delta}_{H}^{-1}\left(x,z\right)+m_{G}^{2}\int_{yw}\delta\Delta_{dN}\left(y,w\right)\tilde{\Delta}_{H}^{-1}\left(w,z\right).
\end{align}
Again the last term vanishes and
\begin{align}
0 & =\int_{y}\tilde{\Delta}_{H}^{-1}\left(z,y\right)\delta\varphi_{d}\left(y\right)+J_{d}\left(z\right)+\int_{x}\delta\varphi_{d}\left(x\right)\tilde{\Delta}_{H}^{-1}\left(x,z\right)\nonumber \\
 & =2\int_{y}\tilde{\Delta}_{H}^{-1}\left(z,y\right)\delta\varphi_{d}\left(y\right)+J_{d}\left(z\right).
\end{align}
The second line follows by the Hermiticity of $\tilde{\Delta}_{H}^{-1}$.
Finally we obtain another wave equation
\begin{equation}
\int_{y}\tilde{\Delta}_{H}^{-1}\left(z,y\right)\delta\varphi_{d}\left(y\right)=-\frac{1}{2}J_{d}\left(z\right).
\end{equation}
Again $\delta\varphi_{d}$ is over-determined, but by a \emph{different} equation this time.

\section{Mechanical Analogy\label{sec:Mechanical-Analogy}}

Here we investigate a very simple mechanical system which illustrates several of the unusual features of the constraint procedure and linear response formulation we have used.
It shows: (a) why the Lagrange multiplier diverges, (b) why constraints must be imposed in the linear response approximation to begin with, (c) why secondary constraints arise.
Consider a unit mass classical particle constrained to move without friction on a circular hoop of radius $r$ in the $x-y$ plane.
Its Lagrangian is
\begin{align}
L & =\frac{1}{2}\dot{x}^{2}+\frac{1}{2}\dot{y}^{2}-\lambda W+j_{x}x+j_{y}y,\\
W & =\frac{1}{4}\left(x^{2}+y^{2}-r^{2}\right)^{2},
\end{align}
where $\lambda$ is the Lagrange multiplier and the form of the constraint $W=0$ is chosen to mimic the singular constraint procedure.
The equations of motion are
\begin{align}
\ddot{x} & =-\lambda\partial_{x}W+j_{x}\nonumber \\
 & =-\lambda\left(x^{2}+y^{2}-r^{2}\right)x+j_{x},\\
\ddot{y} & =-\lambda\partial_{y}W+j_{y}\nonumber \\
 & =-\lambda\left(x^{2}+y^{2}-r^{2}\right)y+j_{y}.
\end{align}
We now consider the source free case $j_{x}=j_{y}=0$.
The constraint terms vanish unless $\lambda\to\infty$ as $x^{2}+y^{2}-r^{2}\to0$.
We set $x^{2}+y^{2}-r^{2}=\eta$ and $\lambda\eta=\omega^{2}$ and take the limit such that $\omega^{2}$ is a constant.
Note that $W=\eta^{2}/4$.
Then the equations of motion become
\begin{align}
\ddot{x} & =-\omega^{2}x,\\
\ddot{y} & =-\omega^{2}y,
\end{align}
with the solutions
\begin{align}
x & =r\cos\left[\omega\left(t-t_{0}\right)\right],\\
y & =r\sin\left[\omega\left(t-t_{0}\right)\right],
\end{align}
where $t_{0}$ and $\omega$ are determined by the initial conditions.
We take as the static solution $x=r$ and $y=0$, which determines $\omega=0$ and $t_{0}=0$.

Now we turn on the sources $j_{x}$ and $j_{y}$ and investigate the linear response by setting $x\to\tilde{x}+\delta x$, $y\to\tilde{y}+\delta y$, $\lambda\to\tilde{\lambda}+\delta\lambda$ where the tilde variables are the source free solutions.
The variation of the constraint is
\begin{equation}
\delta W=\eta\left(\tilde{x}\delta x+\tilde{y}\delta y\right)\to0,
\end{equation}
regardless of the behavior of $\delta x$ and $\delta y$, so long as they are non-singular in the $\eta\to0$ limit.
However, the first order equations of motion become
\begin{align}
\delta\ddot{x} & =-\delta\lambda\eta\tilde{x}-\tilde{\lambda}2\left(\tilde{x}\delta x+\tilde{y}\delta y\right)\tilde{x}-\tilde{\lambda}\eta\delta x+j_{x}\nonumber \\
 & =F_{x}^{\mathrm{rad}}-\omega^{2}\delta x+j_{x}=-\omega^{2}\delta x+j_{x}^{\perp},\\
\delta\ddot{y} & =-\delta\lambda\eta\tilde{y}-\tilde{\lambda}2\left(\tilde{x}\delta x+\tilde{y}\delta y\right)\tilde{y}-\tilde{\lambda}\eta\delta y+j_{y}\nonumber \\
 & =F_{y}^{\mathrm{rad}}-\omega^{2}\delta y+j_{y}=-\omega^{2}\delta y+j_{y}^{\perp},
\end{align}
where we introduce the radial force $\mathbf{F}^{\mathrm{rad}}=-\left[\delta\lambda\eta+\tilde{\lambda}2\left(\tilde{x}\delta x+\tilde{y}\delta y\right)\right]\left(\tilde{x},\tilde{y}\right)$, whose physical function is to balance the applied force normal to the constraint surface, resulting in the net transverse source $\mathbf{j}^{\perp}$.

Now notice the terms proportional to $\tilde{\lambda}\left(\tilde{x}\delta x+\tilde{y}\delta y\right)$ in the equations of motion.
In order for these terms to be well behaved in the limit $\tilde{\lambda}\to\infty$ we must have $\tilde{x}\delta x+\tilde{y}\delta y\to0$, i.e. the response remains within the constraint surface (to first order).
Thus the vanishing of these terms in addition to the vanishing of $\delta W$ is required to fully enforce that the response be tangential to the constraint surface.
We also note that by examining the $\delta\lambda$ terms in the equation of motion one can identify which component of the applied force acts normal to the constraint surface (and hence produce no physical response).

Applying the static solution we find $\mathbf{F}^{\mathrm{rad}}=-\left[\delta\lambda\eta+\tilde{\lambda}2r\delta x\right]\left(r,0\right)$.
For this to be well behaved as $\tilde{\lambda}\to\infty$ requires $\delta x=0$, which also determines $j_{x}^{\perp}=0$ via the $\delta x$ equation of motion.
The $\delta y$ equation of motion is
\begin{align}
\delta\ddot{y} & =j_{y},
\end{align}
with the solution (taking into account the initial conditions $\delta y_{0}=\delta\dot{y}_{0}=0$):
\begin{equation}
\delta y=\int_{0}^{t}\int_{0}^{\tau}j_{y}\left(\tau'\right)\mathrm{d}\tau'\mathrm{d}\tau=\int_{0}^{t}\left(t-\tau\right)j_{y}\left(\tau\right)\mathrm{d}\tau.
\end{equation}

Now we can compare this to the exact solution.
Substituting the ansatz
$x=r\cos\theta\left(t\right)$, $y=r\sin\theta\left(t\right)$, the
Lagrangian and equation of motion become
\begin{align}
L & =\frac{1}{2}r^{2}\dot{\theta}^{2}+j_{x}r\cos\theta+j_{y}r\sin\theta,\\
\ddot{\theta} & =-\frac{j_{x}}{r}\sin\theta+\frac{j_{y}}{r}\cos\theta=\frac{j_{\theta}}{r},
\end{align}
where $j_{\theta}=-j_{x}\sin\theta+j_{y}\cos\theta$ is the tangential component of the force.
The solution satisfying the initial conditions $\theta_{0}=\dot{\theta}_{0}=0$ is
\begin{equation}
\theta=\int_{0}^{t}\left(t-\tau\right)\frac{j_{\theta}\left(\tau\right)}{r}\mathrm{d}\tau.
\end{equation}
To linear order in $j_{\theta}$ the $x$ and $y$ components are just
\begin{align}
x & =r\cos\theta=0,\\
y & =r\sin\theta=\int_{0}^{t}\left(t-\tau\right)j_{\theta}\left(\tau\right)\mathrm{d}\tau,
\end{align}
which, on putting $j_{\theta}=j_{y}+\mathcal{O}\left(j_{x,y}^{2}\right)$, gives
\begin{align}
y & =\int_{0}^{t}\left(t-\tau\right)j_{y}\left(\tau\right)\mathrm{d}\tau,
\end{align}
which is just the solution obtained previously.

If in contrast we never imposed the constraints on the linear response solution we would have the equations of motion
\begin{align}
\delta\ddot{x} & =j_{x},\\
\delta\ddot{y} & =j_{y},
\end{align}
with the solutions
\begin{align}
\delta x & =\int_{0}^{t}\left(t-\tau\right)j_{x}\left(\tau\right)\mathrm{d}\tau,\\
\delta y & =\int_{0}^{t}\left(t-\tau\right)j_{y}\left(\tau\right)\mathrm{d}\tau.
\end{align}
The $\delta y$ component is correct but $\delta x$ is in error already at linear order.
In fact, the solution should not even depend on $j_{x}$ until second order.

%%%%%%%%%%%%%%% START BIBLIOGRAPHY

%\bibliography{refs}

%merlin.mbs apsrev4-1.bst 2010-07-25 4.21a (PWD, AO, DPC) hacked
%Control: key (0)
%Control: author (0) dotless jnrlst
%Control: editor formatted (1) identically to author
%Control: production of article title (0) allowed
%Control: page (1) range
%Control: year (0) verbatim
%Control: production of eprint (0) enabled
%

%%%%%%%%%%%%%%% END BIBLIOGRAPHY

\end{document}